\documentclass[aps,pre,twocolumn,superscriptaddress]{revtex4-1}
\pdfoutput=1
\usepackage{amsmath}
\usepackage{graphicx}
\usepackage{longtable}
\usepackage{color}

\begin{document}

\title{Role of DNA binding sites and slow unbinding kinetics \\ 
in titration-based oscillators}

\author{Sargis Karapetyan}
\affiliation{Department of Physics, Duke University, Durham, NC 27708}
\affiliation{Center for Genomic \& Computational Biology, Durham, NC 27710}

\author{Nicolas E. Buchler}
\affiliation{Department of Physics, Duke University, Durham, NC 27708}
\affiliation{Center for Genomic \& Computational Biology, Durham, NC 27710}
\affiliation{Department of Biology, Duke University, Durham, NC 27708}

\date{\today}

\begin{abstract}
Genetic oscillators, such as circadian clocks, are constantly perturbed by
 molecular noise arising from the small number of molecules involved in
gene regulation. One of the strongest sources of stochasticity is the binary
noise that arises from the binding of a regulatory protein to a promoter in
the chromosomal DNA. In this study, we focus on two minimal oscillators
based on activator titration and repressor titration to understand the key
parameters that are important for oscillations and for overcoming
binary noise. We show that the rate of unbinding from the DNA, despite 
traditionally being considered a fast parameter, needs to be slow to broaden 
the space of oscillatory solutions. The addition of multiple, independent DNA 
binding sites further expands the oscillatory parameter space for the 
repressor-titration oscillator and lengthens the period of both oscillators. 
This effect is a combination of increased effective delay of the unbinding 
kinetics due to multiple binding sites and increased promoter ultrasensitivity 
that is specific for repression. We then use stochastic simulation to show that 
multiple binding sites increase the coherence of oscillations by mitigating 
the binary noise. Slow values of DNA unbinding rate are also effective in 
alleviating molecular noise due to the increased distance from the bifurcation 
point. Our work demonstrates how the number of DNA binding sites and slow 
unbinding kinetics, which are often omitted in biophysical 
models of gene circuits, can have a significant impact on the temporal 
and stochastic dynamics of genetic oscillators. 
\end{abstract}

\maketitle

\section{Introduction}
Genetic oscillatory networks are ubiquitous in nature and perform important 
functions. For example, the cell cycle oscillator regulates cell growth and 
division, whereas the circadian clock regulates the behavior of organisms 
with respect to daily changes in light. These genetic oscillators are used by 
living systems to reliably coordinate various periodic internal processes with 
each other as well as with their rhythmic environment. However, this presents 
a challenge at the cellular level because oscillators have to maintain proper 
timing (temporal coherence of oscillation) in the presence of stochastic 
noise that arises from the small number of regulatory molecules 
in cells \cite{Schrodinger:1944:WLP}.

\begin{figure}[b]
\includegraphics[width=.50\textwidth]{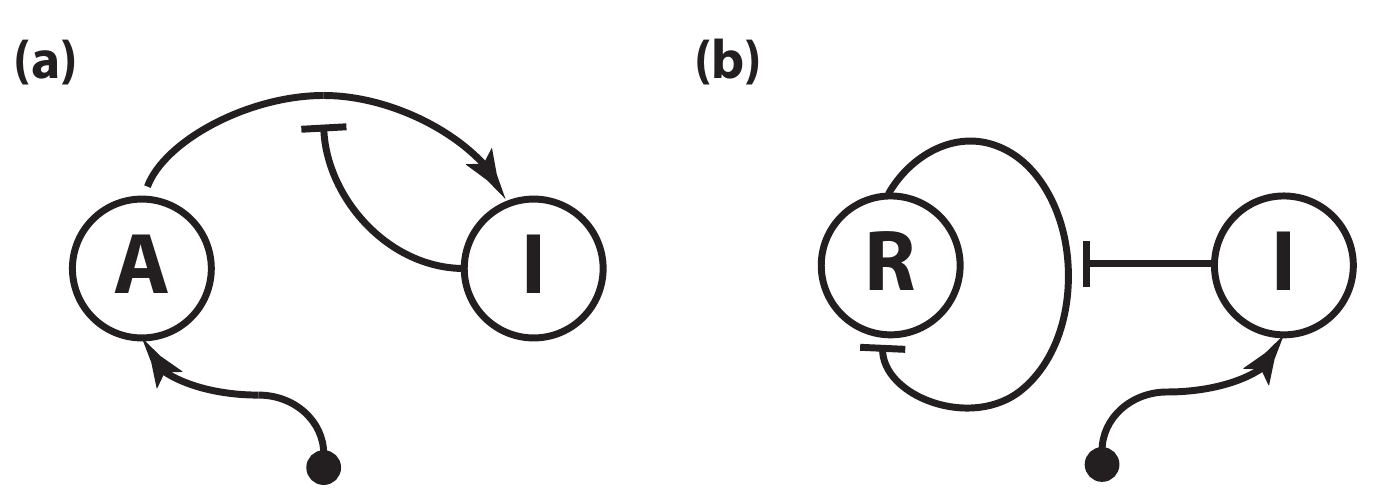}
\caption{(a) In the Activator-Titration Circuit (ATC), the activator is 
constitutively produced at a constant rate and activates the expression 
of the inhibitor, which, in turn, titrates the activator into inactive 
complex. (b) In the Repressor-Titration Circuit (RTC) the constitutively 
expressed inhibitor titrates the self-repressing repressor.}
\end{figure}

A simple mechanism to mitigate the effect of molecular noise would be to 
increase the number of molecules of each species 
\cite{Gonze2002dvs,Forger:2005ej,Gonze2006c}. 
While the number of RNAs and proteins 
made per gene can be large, most cells are fundamentally constrained to 1-2 
gene copies and are subject to binary noise in the first step of gene regulation 
(i.e., transcription factor binding to DNA) \cite{Golding:2005dm,Cai:2006cp}. 
This binary gene regulation noise manifests itself as a 
stochastic temporal pattern of all-or-none
gene activity depending on whether the promoter is bound by the regulatory 
protein or not. Recent work shows that slow DNA binding/unbinding kinetics 
(also called the non-adiabatic limit) can exacerbate the binary noise and 
have profound consequences on gene expression \cite{Hornos:2005ct}, 
epigenetic switching \cite{Walczak:2005ds}, and oscillation 
\cite{Forger:2005ej,Gonze2006c,Feng20121001,Potoyan11022014}. 
Faster kinetic rates and complex gene promoter architectures have 
been proposed as a way to suppress the effect of this binary noise. 
For example, increasing the DNA binding/unbinding rate can increase 
temporal coherence of oscillations via more precise sampling of the 
concentration of transcription factors 
\cite{Forger:2005ej,Feng20121001,Labavic:2013gu,Potoyan11022014} 
or by increasing the distance from a bifurcation point 
\cite{Gonze:2004fv,Gonze2006c}. 
However, transcription factors often have slow DNA unbinding rates 
\cite{FEBS:FEBS1897,GTC:GTC711,Okahata1998,Geertz09102012,Hammar:2014hr}, 
which suggests that these mechanisms are not generally applicable. The 
cooperative binding of a transcription factor to multiple binding sites has also 
been shown to increase temporal coherence of oscillations \cite{Gonze:2002im}. 
However, multiple binding sites do not always lead to cooperativity and 
transcription factor binding to a single DNA site may often be enough to 
effectively activate or repress transcription. \par

To better understand the potential mechanisms that suppress the binary gene 
regulation noise, in particular the influence of slow DNA unbinding rates and 
multiple binding sites, we study an Activator-Titration Circuit (ATC) that has 
been theoretically shown to oscillate \cite{Francois:2005tr}. The ATC consists 
of a constitutively-expressed activator that promotes the expression of the 
inhibitor, which then titrates the activator into an inactive heterodimer 
complex (Fig. 1). Studying the ATC has two advantages. First, it lies at the 
core of animal circadian clocks \cite{Menet:2010wb} and oscillatory 
NF-$\kappa$B signaling \cite{Hoffmann:2002kp,Nelson:2004jna}
and has served as a model of natural genetic oscillators 
\cite{Barkai2000,Vilar:2002ir, Francois:2005tr, Krishna:2006hu, 
Kim:2012cs, Potoyan11022014}. 
Second, the ATC generates the necessary nonlinearities through protein 
titration \cite{Buchler:2008bh} and does not require cooperative binding of 
activator to the inhibitor promoter. Thus, by studying a titration-based 
oscillator, we can better explore the kinetic effects of multiple binding sites on 
coherence independently of the effects that might arise from cooperativity. 
To obtain general insights that are not specific to activation, we also study 
a Repressor-Titration Circuit (RTC), which consists of a self-repressor 
and a constitutively-expressed inhibitor (Fig. 1). This novel titration-based 
oscillator is analogous to the ATC but uses repression instead of 
activation for the transcriptional regulation.

We first characterize these oscillators and how they depend on several 
key parameters in Section II. We deliberately constrain ourselves 
to physiological parameters found in a simple eukaryote 
\textit{S. cerevisiae}, commonly known as budding yeast. 
We show that, in addition to slow mRNA degradation, slow DNA unbinding 
rates of transcription factors are important for providing the necessary delay 
in the negative feedback loop for oscillatory solutions. Thus, both the DNA 
unbinding rate and mRNA degradation rate can set the period of oscillation. 
We then demonstrate that the addition of multiple, independent binding 
sites has nontrivial effects on the ATC and the RTC. While multiple binding 
sites lengthen the period of both oscillators due to an effective increase in 
the delay of negative feedback, they dramatically increase the oscillatory 
solution space of the RTC only. This is because multiple, independent 
binding sites generate ultrasensitivity (i.e. strong nonlinear response to 
changes in regulatory protein concentration) in repression-based 
promoters only, and thus only RTC can benefit from this effect. In section III, 
we use stochastic Gillespie simulations to understand the extent to which 
DNA unbinding rates and numbers of binding sites suppress the 
molecular noise in ATC and RTC oscillators. We show that multiple 
binding sites increase the temporal coherence of oscillations by alleviating 
the binary noise resulting from discrete gene states. We also show that 
slower values of DNA unbinding rates are best for coherent oscillations in 
simple titration-based oscillations. Last, we compare and contrast our 
results on temporal coherence with those of previous models of 
genetic oscillators in Section IV.

\section{Biophysical model of ATC and RTC}
Oscillators require negative feedback with nonlinearity and time delays 
\cite{Novak2008}. Mechanistically, negative feedback on gene expression 
can occur transcriptionally via repressors 
\cite{Elowitz:2000gg,Atkinson2003597,Stricker:2008jq} or post-transcriptionally 
via phosphorylation \cite{Novak:1993fj,Nakajima2010898,Yang:2013gy}, degradation 
\cite{Novak:1993fj,Wong:2007hz,Yang:2013gy}, or titration of activators into 
inactive complexes by inhibitors \cite{Barkai2000,Vilar:2002ir, Francois:2005tr,
Krishna:2006hu,Kim:2012cs,Potoyan11022014}. The ATC is a 
minimal two-gene circuit that can oscillate through the use of protein 
titration both as a source of nonlinearity and indirect negative feedback. 
In the first phase of oscillation, high levels of free activator homo-dimerize, 
bind the promoter, and overproduce inhibitor until all free activator has 
been titrated into inactive heterodimer. In the second phase of 
oscillation, the remaining activator will unbind from the inhibitor promoter 
and be sequestered by inhibitor, thus causing the promoter to return to 
low levels of expression of inhibitor. The levels of inhibitor will 
eventually decline to a point where free activator can re-accumulate 
and restart the cycle.

\begin{figure}[b]
\includegraphics[width=.50\textwidth]{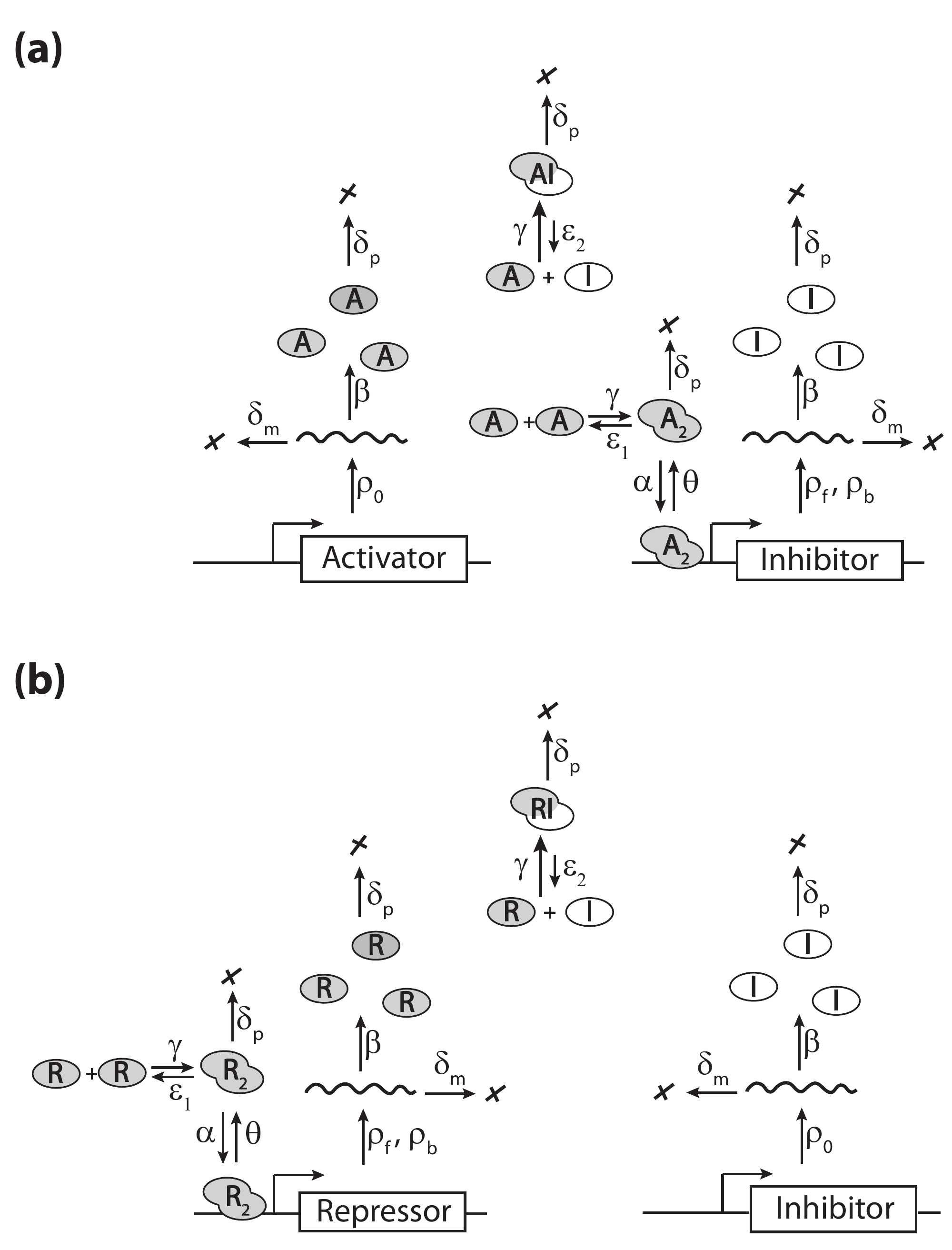}
\caption{A biophysical model for ATC (a) and RTC (b) with 
explicit transcription, translation, protein-protein and protein-DNA
interactions. Each arrow corresponds to a reaction rate in Eqs. 1-3. 
Neither of these titration-based oscillators have been 
built or studied by synthetic biologists.}
\end{figure}

In the RTC, protein titration is used exclusively as a source of 
nonlinearity and the negative feedback is directly achieved through 
auto-repression. In the first phase of oscillation, high levels of free 
repressor will homo-dimerize, bind directly to its own promoter, and 
repress its transcription. The free repressor will be 
titrated away by the constitutively expressed 
inhibitor. In the second phase of oscillation when free repressor levels are 
low, the remaining repressor will unbind from the promoter, returning to 
high levels of transcription and the rapid over-production of free repressor. 
As we will show below, the indirect versus direct nature of negative 
feedback in ATC and RTC is responsible for many of the differences 
between these two titration-based oscillators.\par

\subsection{ATC and RTC oscillators with a single DNA binding site}
Even simple genetic circuits such as the ATC and RTC include many 
reactions and parameters (Fig. 2). An exhaustive search over all the 
parameter space was not feasible, and we decided to constrain our 
parameter space by studying synthetic gene circuits that could be built 
in budding yeast. Synthetic genetic oscillators have been useful tools to 
understand the properties of natural oscillators. For example, a 
synthetic oscillator built in bacteria \cite{MondragonPalomino:2011it} was 
useful in understanding entrainment capabilities of genetic oscillators, as 
well as elucidating sources of stochasticity that affected entrainment. 
Surprisingly, all synthetic genetic oscillators built to date have neglected 
protein titration, a common mechanism in natural oscillators. 
To this end, we built a mathematical model of ATC and RTC 
oscillators using a basic leucine zipper (bZIP) transcription factor 
that dimerizes and binds DNA, and a rationally-designed inhibitor that 
binds free bZIP into an inactive heterodimer. These synthetic components 
have been successfully used in yeast \cite{Buchler:2009fz} and, importantly, 
many of the protein-protein and protein-DNA binding affinities of this bZIP and 
inhibitor pair are known \cite{Krylov:1995wx,Cao01091991}; see Table I. 
We fixed these parameters and scanned through a range of other biophysical 
parameters to understand which ones affect oscillation. Our results 
should help guide future experimental implementation of synthetic ATC 
and RTC oscillators in yeast. \par

\begin{figure*}
\includegraphics[width=1.0\textwidth]{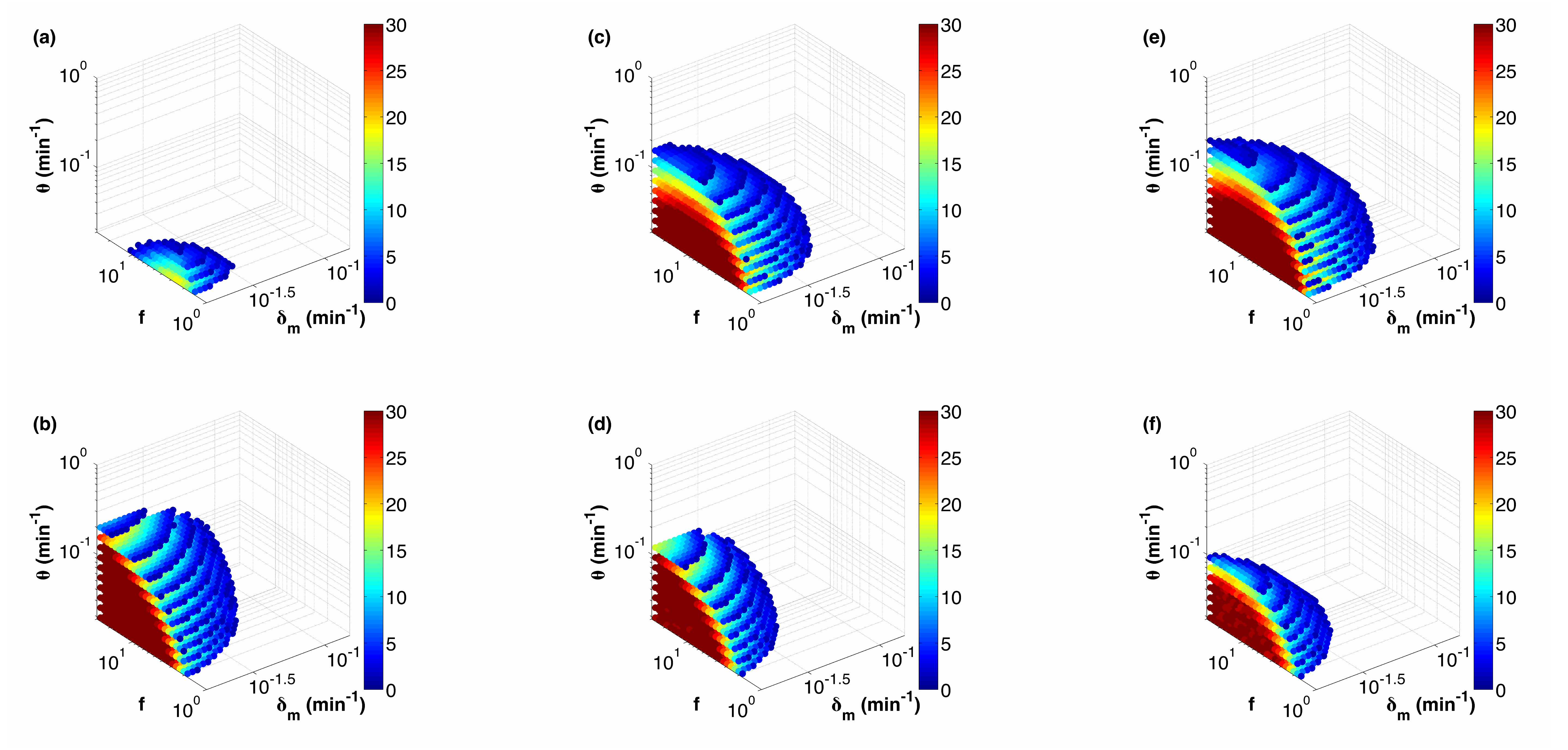}
\caption{(Color online) Parameter space of oscillatory solutions on a 
logarithmic scale for RTC (top) and ATC (bottom) with increasing DNA 
binding sites. The colormap shows the number of $\rho_f$ values that exhibited 
oscillations for each combination of $f$, $\delta_m$, and $\theta$. (a) and (b), 
single DNA binding site for RTC and ATC, (c) and (d) three independent binding 
sites for RTC and ATC, (e) and (f), three synergistic binding sites for RTC 
and ATC, where activation/repression strength is $f^2$ when more 
than one activator/repressor is bound.}
\end{figure*}

The biophysical model of our ATC and RTC circuits is based on chemical 
mass-action kinetics where the dynamic variables are the mean 
concentrations of all molecular species. The ODEs that correspond to 
the reactions in Fig. 2 are the following:

\begin{widetext}
\begin{subequations}
\allowdisplaybreaks[3]
\begin{align}
\frac{d[G_0]}{dt}&=-{\alpha}[G_0][X_2]+{\theta}[G_1] \\
\frac{d[G_1]}{dt}&={\alpha}[G_0][X_2]-{\theta}[G_1]\\
\frac{d[I]}{dt}&=\beta[r_I]-\gamma[X][I]+\epsilon_2[XI]-\delta_p[I]\\
\frac{d[X]}{dt}&=\beta[r_X]-\gamma[X][I]+\epsilon_2[XI]-2\gamma[X]^2
+2\epsilon_1[X_2]-\delta_p[X]\\
\frac{d[XI]}{dt}&=\gamma[X][I]-\epsilon_2[XI]-\delta_p[XI]\\
\frac{d[X_2]}{dt}&=\gamma[X]^2-\epsilon_1[X_2]-\delta_p[X_2]-{\alpha}[G_0]
[X_2]+{\theta}[G_1]
\end{align}
\end{subequations}
\end{widetext}

With $[r_X]$ and $[r_I]$ described by:

\begin{subequations}
	\begin{align}
		\frac{d[r_X]}{dt}&={\rho_0}[G_T]-\delta_m[r_X] \\
		\frac{d[r_I]}{dt}&={\rho_f}[G_0]+{\rho_b}([G_T]-[G_0])-{\delta_m}[r_I]
	\end{align}
\end{subequations}

for the ATC, where X=A (activator), and	

\begin{subequations}
\begin{align}
\frac{d[r_X]}{dt}&={\rho_f}[G_0]+{\rho_b}([G_T]-[G_0])-{\delta_m}[r_X]\\
\frac{d[r_I]}{dt}&={\rho_0}[G_T]-\delta_m[r_I]
\end{align}
\end {subequations}

for the RTC, where X=R (repressor). The first two equations represent the 
dynamics of promoter DNA where $[G_0]$ and $[G_1]$ are the mean 
concentrations of free and bound promoter, respectively. The molar 
concentration of total DNA $[G_T] =[G_0]+[G_1] =1/(N_A\cdot V)=1/24$ ${\rm nM}$ 
where $N_A$ is the Avogadro constant and $V$ is the yeast cell volume; 
see Table 1. Here, we consider only a single DNA binding site, but we will 
later expand our analysis to include multiple binding sites. At any instant, 
the promoter is either free or bound. The probability of free or bound 
promoters is equal to the ratio of concentrations 
$[G_0] /[G_T]$ or $[G_1]/[G_T]$, respectively. The other 
equations describe the mean concentration dynamics of the respective molecular 
species such as mRNA ($r_I$, $r_X$), monomeric protein ($I$ or $X$), and 
dimeric proteins ($X_2$, $XI$), where $X$ stands for the activator $A$ or 
repressor $R$, respectively. The regulatory homodimer $X_2$ associates with 
$G_0$ at a rate $\alpha$ to form $G_1$, which dissociates at the rate $\theta$. 
The $r_I$ and $r_X$ are the inhibitor and activator/repressor mRNAs. For the 
ATC, the activator mRNA $[r_X]$ is transcribed constitutively at the rate 
$\rho_0$, where as the inhibitor mRNA is transcribed at rates $\rho_f$ and 
$\rho_b$ from free and bound DNA, respectively (Eqs. 2a,b). In contrast, for 
the RTC, the repressor mRNA is transcribed at rates $\rho_f$ and $\rho_b$ 
(Eqs. 3a,b) while inhibitor mRNA $r_I$ is constitutively transcribed at the 
rate $\rho_0$. We assume that all mRNA species are degraded at the same 
rate $\delta_m$ and translated into proteins with the same rate $\beta$. 
The activator/repressor $X$ protein dimerizes into active homodimer 
$X_2$ and forms inactive heterodimer $XI$ with the inhibitor protein 
$I$ at the same rate $\gamma$. The homodimer and heterodimer 
dissociation rates are $\epsilon_1$ and $\epsilon_2$, respectively. 
We assume that all protein species are stable and diluted by cell 
growth at rate $\delta_p$. \par

\subsection{DNA unbinding kinetics influence oscillation}

Our parameters were restricted to physiological values from yeast (see 
Table I). Most parameters were kept fixed, but we varied four key parameters. 
The first parameter was the mRNA production rate ($\rho_f$) of free, unbound 
promoter because a desired expression level can easily be selected from 
existing promoter libraries \cite{Keren2013MSB}. Second, we varied the 
activation/repression strength ($f$), which is the ratio of the larger $\rho$ divided by the 
smaller $\rho$. Thus, $f = \rho_b / \rho_f$ for the ATC and $f= \rho_f / \rho_b$ for the RTC. The ratio $f$ can 
be tuned by appropriate choice of activation or repression domains fused to our bZIP transcription factor
\cite{Baron:1997uf,Belli:1998vt,PerezPinera:2013jt}.
The third parameter was the mRNA degradation rate ($\delta_m$), which is 
known to set the time scale of the ATC oscillator \cite{Francois:2005tr}. Last, 
we varied the DNA unbinding rate ($\theta$) because it is our point of focus 
and this parameter can vary between different transcription factors. 
The DNA dissociation constant ($K_d$) fixes the DNA binding rate 
$\alpha = \frac{\theta}{K_d}$; see Appendix for details. \par

We divided the physiological range of each variable parameter into 30 
values (on a logarithmic scale) and evaluated the long-term dynamics of 
a total of $(30)^4$ parameter sets per circuit. We solved the ODEs over 
time for each set of ($\rho_f$, $f$, $\delta_m$, $\theta$). A solution was 
classified as oscillatory if the trough of activator or repressor homodimer 
concentration was below the $K_d$ of DNA-binding and if the peak was 
above $2K_d$; see Appendix for justification. We noticed that 
$\rho_f$ had the smallest effect on the number of oscillatory 
solutions and, thus, we plot the marginal frequency distribution of oscillatory 
solutions over $f$, $\delta_m$, and $\theta$ in Fig. 3. We see that strong 
activators (large $f$ for the ATC), stable mRNAs (small $\delta_m$), and 
slow DNA unbinding rates (small $\theta$) generally favor oscillation. The 
last two parameters dictate the timescale of the delay in the negative 
feedback loop. Increased delay supports oscillation and, thus, the largest 
number of oscillatory solutions occur at the smallest $\theta$ and $\delta_m$ 
for both RTC (Fig. 3a) and ATC (Fig. 3b). The parameter space of stable 
oscillations is larger in ATC relative to RTC for a single binding site because 
of the additional step and delay in the negative feedback loop: negative 
feedback through the activator in the ATC is indirect (i.e. activator regulates 
the expression of inhibitor, which then inhibits its activity), where as the 
self-repressor in the RTC is direct (i.e. repressor regulates its own expression).\par

\begin{figure}[t]
\includegraphics[width=0.5\textwidth]{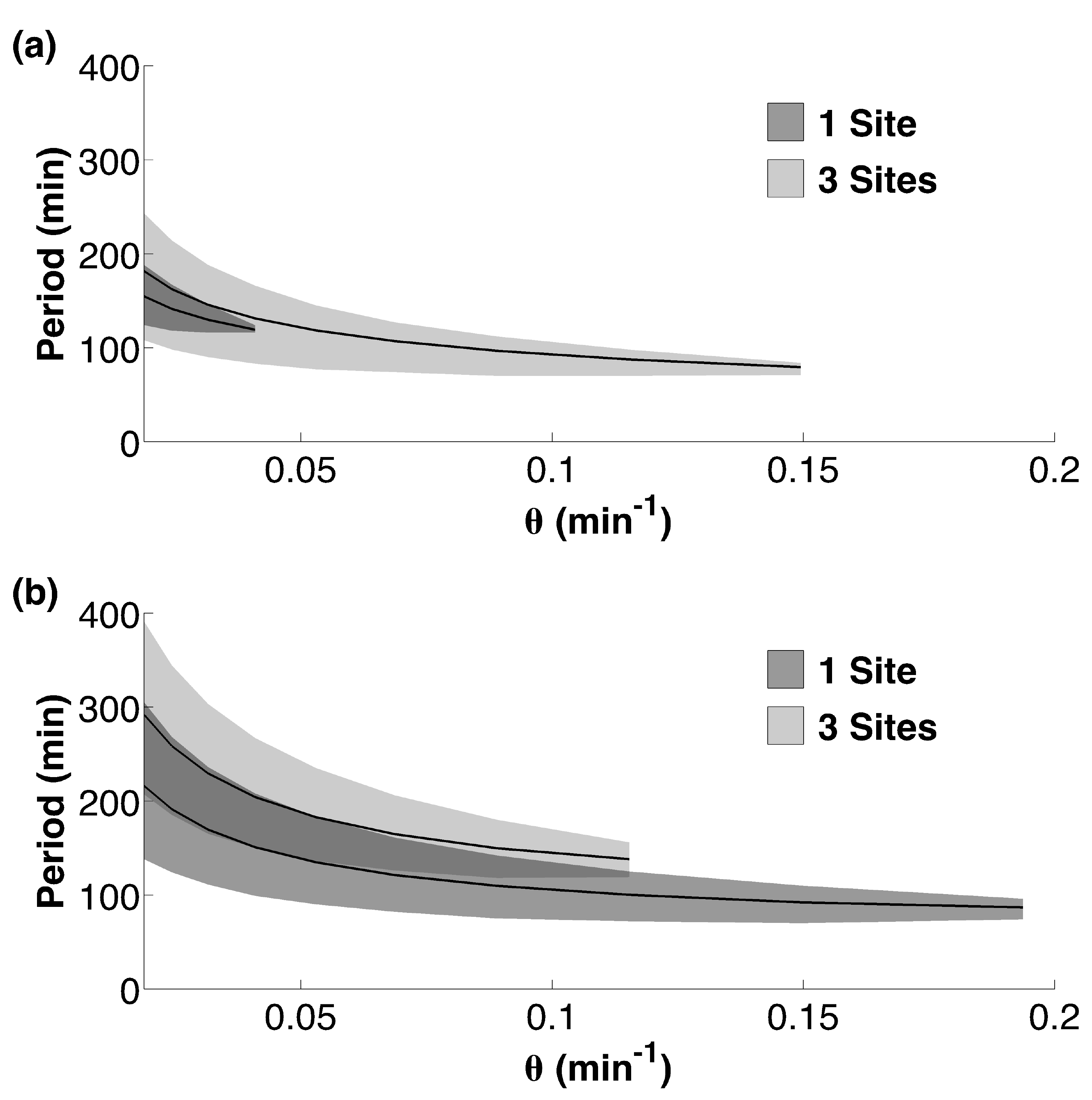}
\caption{The DNA unbinding rate $\theta$ sets the period of the oscillations 
for RTC (a) and ATC (b) at slow unbinding rates. The mean period of 
oscillatory solutions for a given $\theta$ is shown (solid black line) with 
the shaded area representing the range of periods.} 
\end{figure}

The period of oscillation $\tau$ should be set by the timescale of the 
slowest parameters in the delay. The negative feedback in our circuits is 
dominated by DNA unbinding rate $\theta$ and mRNA degradation rate 
$\delta_m$ \cite{Francois:2005tr}. This relationship can be seen in Figure 4 
where the DNA unbinding rate sets the oscillation period at low $\theta$. An 
increase in $\theta$ leads to mRNA degradation rate ($\delta_m$) becoming 
the slower timescale at which point $\tau$ becomes flatter and less 
dependent on $\theta$. Eventually, a bifurcation occurs at a critical, maximum 
value of $\theta_{\rm max}$ which leads to loss of the stable limit cycle. A 
similar relationship exists for the mRNA degradation rate $\delta_m$; 
see Figure S2.

\subsection{Multiple DNA binding sites affect ATC and RTC 
oscillators differently}

\begin{figure}[t]
\includegraphics[width=0.5 \textwidth]{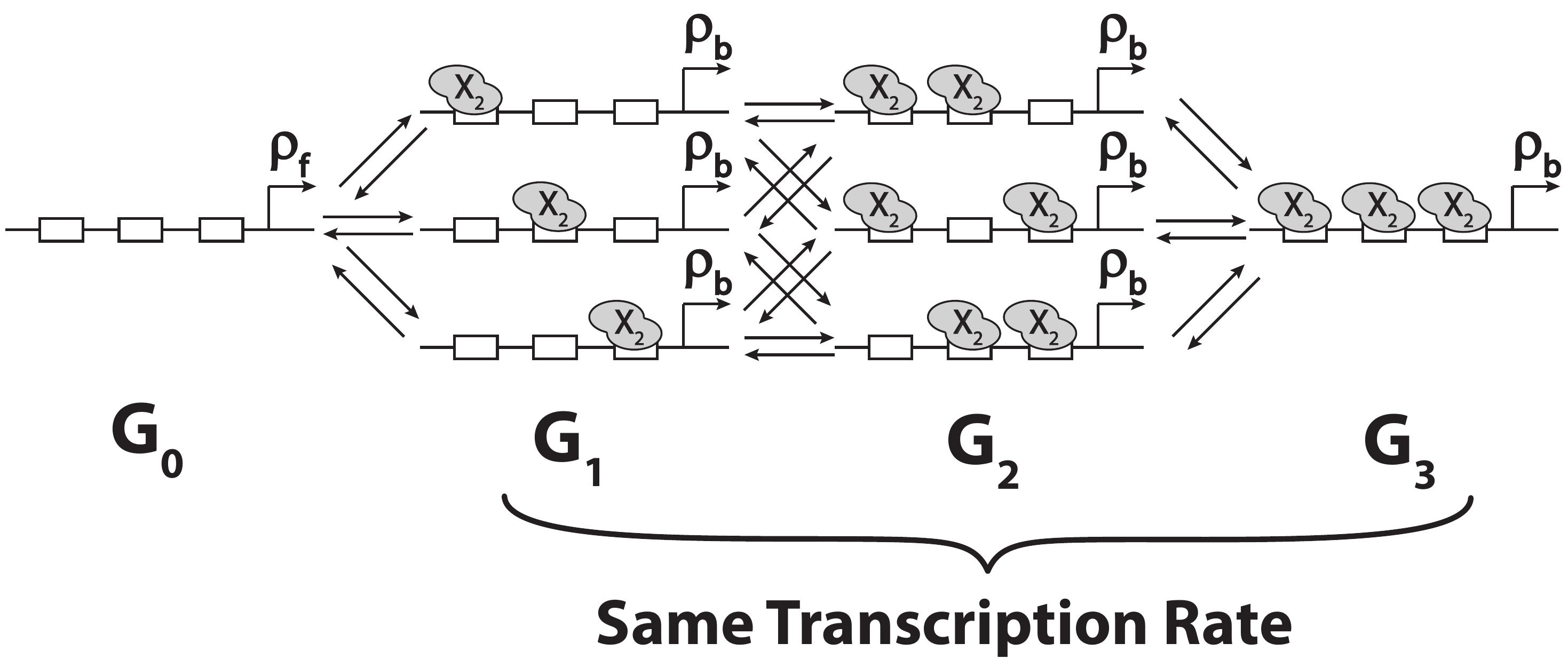}
\caption{Transitions between promoter states for multiple 
DNA binding sites (n=3). $G_i$ denotes the set of promoter states with 
$i$ out of total $n$ binding sites occupied by activator (X=A) or repressor 
(X=R) dimers. There are $n-i$ ways of switching from $G_i$ to $[G_{i+1}]$ 
via the binding of $X_2$, and there are $i$ ways of switching from 
$G_{i}$ to $G_{i-1}$ by the unbinding of $X_2$. Our model conservatively 
assumes that the binding of $X_2$ does not affect the binding or unbinding 
of the next transcription factor to an adjacent site (no cooperativity). 
We also assume that the transcription rate is equal to 
$\rho_b$ for $G_1$, $G_2$, $\ldots G_n$ promoter states.}
\end{figure}

This role of DNA unbinding rate in generating delays led us to 
hypothesize that multiple DNA binding sites should increase the parameter 
space of oscillations and lengthen the period of the oscillators. We 
reasoned that if the occupancy of {\it any} binding site by a transcription 
factor activates or represses transcription, then the effective unbinding 
rate ($\theta_n$) from a state of saturated DNA binding to the unbound 
DNA state ($G_0$, where the transcription rate changes) should decrease 
with the increasing number of binding sites ($n$). We can show that 
$\theta_n = \theta/H_n$, where $H_n$ is the n-th harmonic number 
(see Supplement \cite{Supplement}). \par

The addition of multiple DNA binding sites to our model will modify Eqs. 
(1a-b, 1f) by increasing the number of promoter states that can be 
bound by $X_2$; see Figure 5 and Supplement \cite{Supplement}. 
For three binding sites (n=3), our new Eqs (1a-b) are:

\begin{eqnarray}
\frac{d[G_0]}{dt} &=& -3 \alpha \cdot [G_0] [X_2] + \theta [G_1] \notag \\
\frac{d[G_1]}{dt} &=& 3 \alpha \cdot [G_{0}] [X_2] - (\theta + 
2 \alpha \cdot [X_2]) [G_{1}] + 2 \theta \cdot [G_{2}] \notag \\
\frac{d[G_2]}{dt} &=& 2 \alpha \cdot [G_{1}] [X_2] - (2 \theta + 
\alpha \cdot [X_2]) [G_{2}] + 3 \theta \cdot [G_{3}] \notag \\
\frac{d[G_3]}{dt} &=& \alpha \cdot [G_{2}] [X_2] - 3 \theta \cdot [G_3]
\end{eqnarray}

where the total concentration of DNA $[G_T] = [G_0] + 
[G_1] + \ldots + [G_n]$ is fixed to $1/(N_A\cdot V)=1/24$ ${\rm nM}$. 
For three binding sites (n=3), the term 
$-{\alpha}[G_0][X_2]+{\theta}[G_1]$ in Eq. (1f) is replaced with:

\begin{equation}
-\sum_{i=0}^{3} (3-i) \alpha \cdot [G_i] [X_2] + 
\sum_{i=0}^{3} i \cdot \theta [G_i]
\end{equation}

$G_i$ denotes the promoter states with $i$ out of total $n$ binding 
sites occupied by activator (X=A) or repressor (X=R) dimers. The 
factors in front of each term represents the amount of degeneracy 
of each state, i.e $[G_i]$ has $i$ bound sites, thus $i$ ways of 
switching to $[G_{i-1}]$. Therefore, we have the term $i\cdot \theta [G_i]$. 
At the same time, $[G_i]$ has $n-i$ vacant sites, so it has $n-i$ ways of 
switching to the state $[G_{i+1}]$. Thus, we have the term 
$(n-i) \alpha \cdot [G_{i}] [X_2] $.

The addition of two more independent DNA binding sites dramatically increased 
the oscillatory space of the RTC (Fig. 3c), while slightly decreasing the 
oscillatory space of the ATC (Fig. 3d). These opposite results arise from a 
compound effect. First, two extra binding sites decreased the effective 
unbinding rate for the promoter to be fully vacated by half 
($\theta / H_3 \approx \theta/2$). This decrease in effective $\theta$ 
increased the delay and resulted in some improvement in oscillations for 
both RTC and ATC. This effect is best observed in the increased period of 
both oscillators (Fig. 4). The second, more dominant effect is the 
fundamental difference in how the promoter sensitivity changes with 
multiple, independent binding sites. It is well established that nonlinear 
promoter responses facilitate oscillation \cite{Novak2008}. We use the 
logarithmic sensitivity ($S$) to quantify the nonlinearity in the promoter 
response, where $S = {\rm dlog}P/{\rm dlog}[X_2]$ \cite{Bintu:2005bn}. 
$P$ is the synthesis rate of the promoter and $[X_2]$ is the 
activator/repressor homodimer concentration that regulates the promoter. 
As shown previously \cite{Bintu:2005bn,Lengyel2014}, an increase in the 
number of independent repressive binding sites will increase the 
magnitude of $S$ and create an ultrasensitive promoter response, 
(i.e. $|S|>1$, see Supplement \cite{Supplement}). However, increasing 
the number of independent activating binding sites cannot generate an 
ultrasensitive promoter response ($|S| \le 1$); see Discussion in 
\cite{Bintu:2005bn}. In fact, the logarithmic sensitivity for activation 
actually decreased with the number of binding sites at our physiological 
concentrations (see Supplement \cite{Supplement}). 
This difference is the reason why the RTC and ATC oscillators exhibited 
fundamental differences to increased number of binding sites. Our work 
shows that synthetic repression-based oscillators are preferable designs because 
the RTC gets an effective boost in promoter ultrasensitivity simply by 
adding multiple, independent binding sites. \par

We also tested whether synergistic repression or activation might change 
our results. Synergistic activation or repression occurs when the states that 
have more than one binding site occupied (i.e. $G_2$ and $G_3$) are 
activated/repressed $f^2$-fold instead of $f$-fold because they can interact 
with RNA polymerase at several interfaces \cite{Bintu:2005bn}. Although 
this synergy increased the activation or repression strength, it did not 
significantly change the oscillatory parameter space (Fig. 3e,f).\par

\section{Stochastic Simulations}
Deterministic simulations were useful for understanding how DNA 
unbinding rate and the number of binding sites affect the phase space and 
period of oscillation. However, they cannot provide insights into the loss of 
temporal coherence that arises from stochastic gene expression. To this 
end, we used the Gillespie algorithm \cite{Gillespie:1977ww} to simulate 
stochastic chemical reaction kinetics and investigate how DNA binding/
unbinding dynamics and the addition of binding sites affect the temporal 
coherence of ATC and RTC oscillators. \par

\begin{figure}
\includegraphics[width=.5\textwidth]{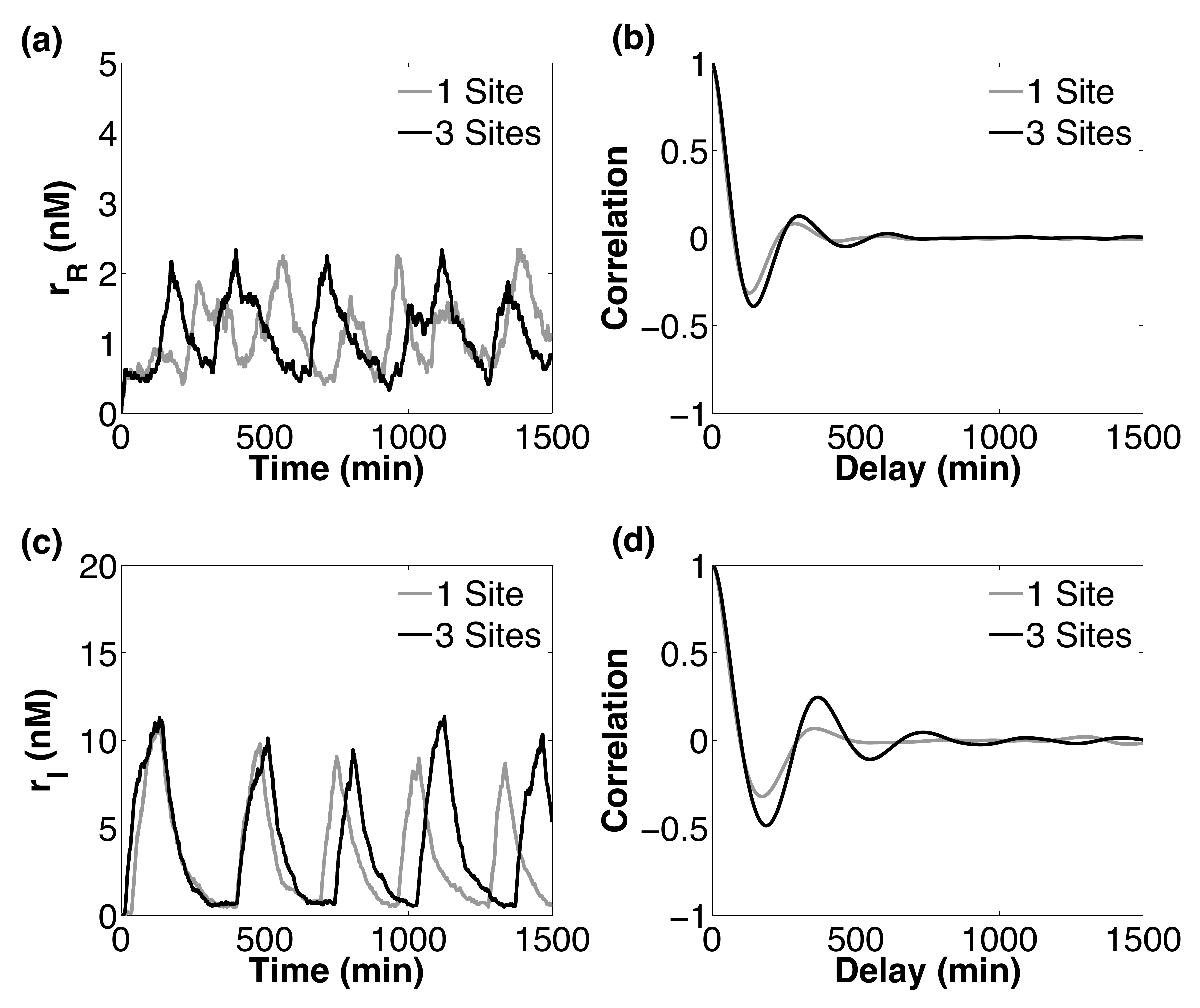}
\caption{Sample stochastic trajectories for one and three binding sites for RTC 
(a) and ATC (c), and their autocorrelation functions (b,d). The variable 
parameter values ($\theta$, $\delta_m$, $\rho_f$, $f$) were fixed to 
(0.02 min$^{-1}$, 0.0159 min$^{-1}$, 0.8928 min$^{-1}$, 3.63) for the RTC 
and (0.02 min$^{-1}$, 0.0186 min$^{-1}$, 0.1781 min$^{-1}$, 30) for the ATC. 
We chose parameters that produced oscillation over the largest 
range of DNA unbinding rates. The rest of the parameters are given in Table 1.}
\end{figure}

For each ATC and RTC, we quantified temporal 
coherence by calculating the autocorrelation function of mRNA transcripts 
levels (repressor mRNA for the RTC and inhibitor mRNA for the ATC); see 
Fig. 6. In the absence of noise, an undamped oscillatory signal will have an 
undamped, periodic autocorrelation function. The presence of noise will 
stochastically perturb period and phase, such that the autocorrelation now 
exhibits dampening or loss of temporal coherence. We quantified the loss 
of coherence by measuring the rate of exponential decay ($e^{-t/\tau_0}$) 
of the envelope of a periodic (${\rm cos}(2 \pi t/\tau)$) autocorrelation 
function (see Appendix for details). Similar to other studies 
\cite{Gonze:2002im, Gonze2006c}, our metric for temporal coherence is the 
normalized autocorrelation function decay rate, which is the ratio of 
timescales $\tau_0/\tau$. A larger ratio indicates better temporal 
coherence. We varied the DNA unbinding rate ($\theta$) and number of 
binding sites ($n$) to understand the role of each feature in resisting 
molecular noise. \par

\subsection{DNA Unbinding Rate}

Our results show that ATC and RTC oscillators with smaller 
DNA unbinding rates exhibit better temporal coherence (Fig. 7).
Lower $\theta$ increases 
the temporal coherence of the oscillations because of the increased 
distance of the dynamical system from the bifurcation point 
($\theta_{\rm max}$); see Figure 8. Eventually there is another bifurcation at small 
$\theta_{\rm min}$, but these unbinding rates are unphysiological and 
do not affect our conclusions regarding biophysical ATC and RTC oscillators. 
Strikingly, some $\theta > \theta_{\rm max}$, which do not show 
deterministic oscillation, exhibit oscillation in the presence of noise. 
This phenomenon is consistent with coherence resonance 
\cite{Pikovsky1997prl} which has been observed in other excitable, 
genetic circuits near oscillatory bifurcation points 
\cite{Vilar:2002ir,Gonze:2004fv}. \par

\subsection{Multiple DNA Binding Sites}

Increasing the number of binding sites ($n$) also increased the temporal 
coherence of ATC and RTC oscillators over all DNA unbinding rates (Fig. 7). 
To better understand this result, we must consider the effect of stochastic 
binding and unbinding of regulators on the variance of gene expression. 
In the phase of changing activator/repressor concentrations, the binding 
sites start being occupied or vacated. Each additional binding site 
introduces an additional DNA binding state. Because we treat the 
expression level of all bound DNA states as equivalent ($\rho_b$), 
the spontaneous binding and unbinding events that occur between states 
that have at least one binding site occupied have no effect on 
transcription; see Fig. 5. These ``protected'' states act as a buffering 
mechanism to mitigate the effects of binary noise on temporal coherence. \par

\begin{figure}[t]
\includegraphics[width=0.5\textwidth]{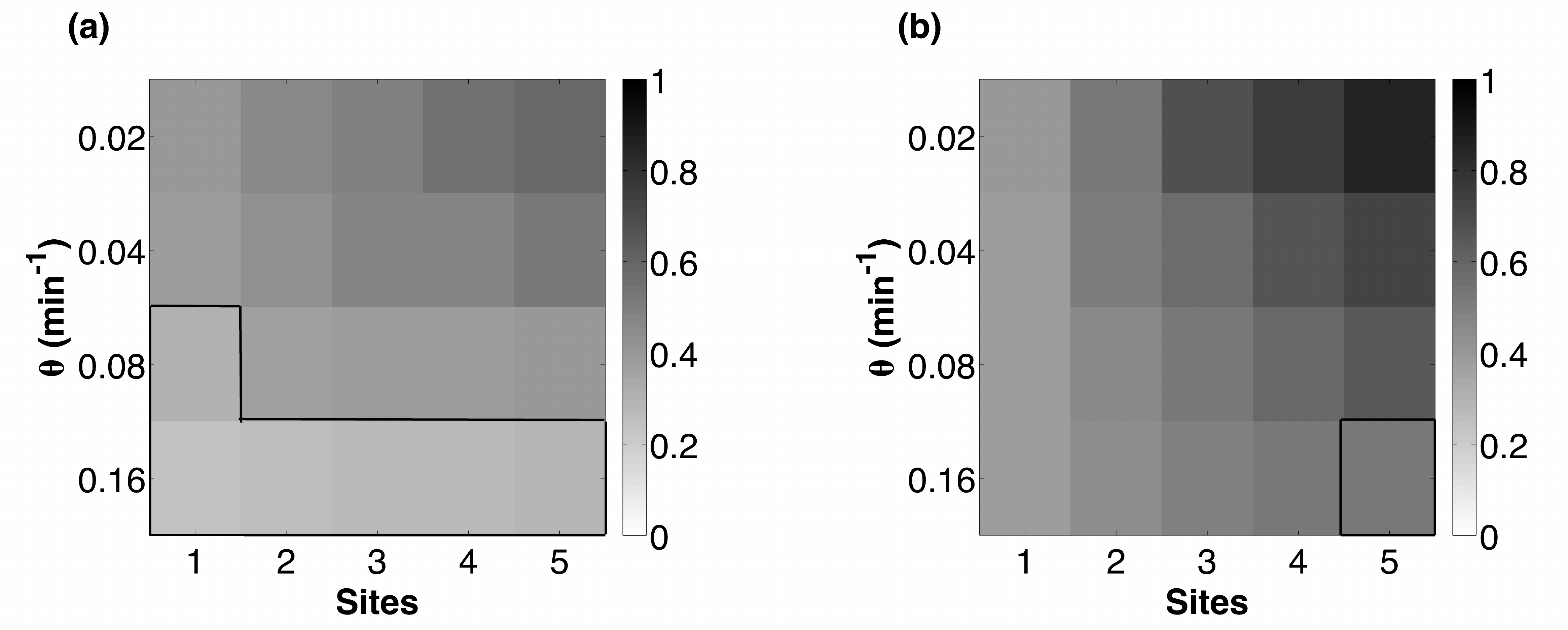}
\caption{The normalized autocorrelation function decay rate 
for the RTC (a) and ATC (b) for varying $\theta$ and number of 
binding sites. All parameters, except $\theta$, are the same 
as in Fig. 6. Boxed, outlined regions are parameters past the bifurcation point 
($\theta_{\rm max}$) where deterministic oscillations are unsustainable and 
damped, yet exhibit stochastic excitable oscillations.}
\end{figure}

\section{Discussion}

We analyzed the properties of two titration based genetic oscillators, the 
activator-titration circuit (ATC) and the repressor-titration circuit (RTC). The 
focus of our study was to understand how the number of DNA binding sites 
and slow unbinding kinetics in promoters mitigate or exacerbate the binary 
gene regulation noise. First, we showed that multiple DNA binding sites 
and slow unbinding kinetics were important for providing the necessary 
delay in the negative feedback loop for oscillatory solutions. The role of 
slow DNA binding/unbinding in providing delay for oscillations is consistent 
with prior work on a small negative feedback oscillator \cite{Feng20121001}. 
Second, we used stochastic simulation to show that slower DNA unbinding 
rates exhibited better temporal coherence, a result which appears at odds 
with previous work on circadian clocks and NF-$\kappa$B oscillators 
\cite{Gonze:2004fv, Forger:2005ej, Potoyan11022014} and which is more 
in line with the results obtained for a small negative feedback oscillator 
model \cite{Feng20121001}. This previous work showed that slower 
DNA unbinding kinetics negatively affected the temporal coherence 
for two reasons. First, slow DNA unbinding increased the stochasticity 
of gene expression due to imprecise concentration sampling, which 
decreased the temporal coherence of oscillations 
\cite{Forger:2005ej,Potoyan11022014}. Second, slower DNA 
unbinding ($\theta$) pushed the dynamical system towards 
$\theta_{\rm min}$ bifurcation point, which made it less robust to noise 
\cite{Gonze:2004fv}. These results are different from ours 
because the delays in the circadian clock and NF-$\kappa$B models 
rely on slow intermediate steps (e.g. phosphorylation and/or nuclear 
transport) in the negative feedback loop. Unlike our titration-based
oscillators in Fig. 8, these models do not have $\theta_{\rm max}$ and 
still oscillate at infinitely fast unbinding rates where the promoter 
dynamics are in steady-state.\par

\begin{figure}[t]
\includegraphics[width=0.4\textwidth]{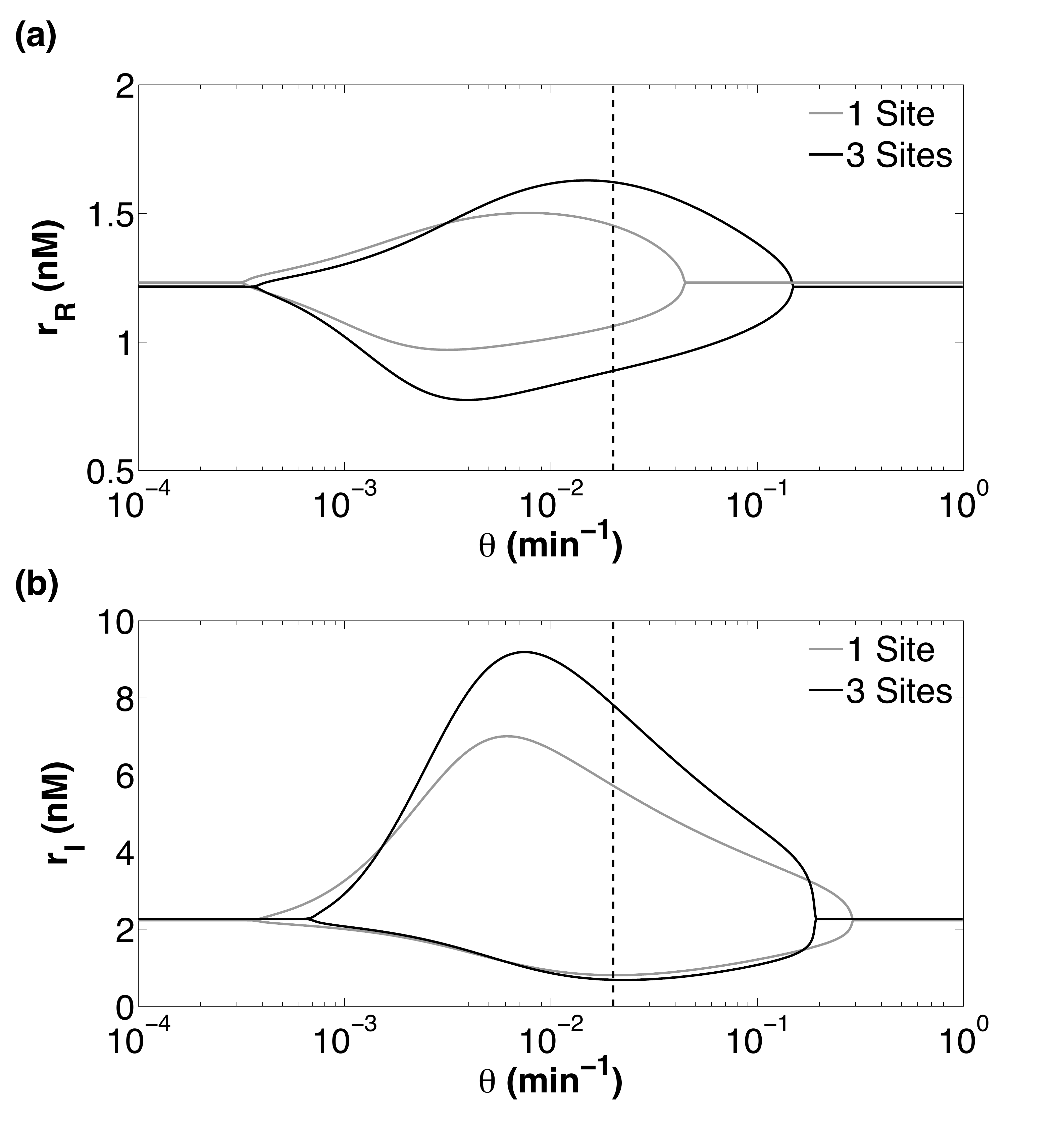}
\caption{Bifurcation diagram of the RTC (a) and ATC (b) oscillators 
as a function of DNA unbinding rate ($\theta$). All parameters, except 
$\theta$, are the same as in Fig. 6. There are two bifurcation points 
($\theta_{\rm max}$, $\theta_{\rm min}$) and the amplitude of mRNA 
oscillation is shown by the upper and lower branches. Physiological 
values of $\theta$ are to the right of the dashed vertical line.}
\end{figure}

We observed the opposite effect for our titration-based oscillators 
because physiological $\theta$ overlaps the $\theta_{\rm max}$ 
bifurcation point for ATC/RTC. Thus, lowering $\theta$ always 
increases the robustness in ATC/RTC because the dynamical system 
is moving away from $\theta_{\rm max}$ and deeper into oscillatory 
parameter space. This phenomenon likely explains the similar results 
presented in \cite{Feng20121001}. The influence of DNA unbinding rate on 
temporal coherence depends on the structure of the underlying bifurcation 
diagram of each oscillator as a function of $\theta$. Changes in topology, 
mechanism, and parameters can change the bifurcation diagram and, 
thus, the influence of DNA unbinding rate on temporal coherence of 
oscillation may also change.\par

Last, we demonstrate that multiple independent binding sites consistently 
increased the temporal coherence of oscillations by alleviating the binary 
noise resulting from binary gene states. Our results agree with previous work, 
which showed that multiple, cooperative DNA binding sites increased the 
coherence of circadian clocks \cite{Gonze:2002im}. However, in contrast to 
our results, the temporal coherence of circadian clocks peaked at 3 binding 
sites and then decreased with additional sites. The difference likely arises 
from our slower DNA binding/unbinding rates, where the ATC and RTC 
oscillators spend significant time in protected states that are buffered 
against molecular noise. In contrast, the circadian clock model spends 
very little time in the intermediate protected states between fully free or 
fully bound promoters and, therefore, the increased coherence is only 
due to cooperativity \cite{Gonze:2002im}. The idea of buffering to reduce 
noise in gene circuits has been discussed in the context of decoy binding 
sites \cite{Burger:2012pre}. However, this requires fast DNA 
binding/unbinding, where as buffering through promoter states requires 
slow DNA binding/unbinding. We note that increased temporal coherence 
due to protected states is a stochastic effect because the addition of 
binding sites consistently increased the coherence of ATC oscillators, 
despite occasionally pushing it past the bifurcation point at 
$\theta_{\rm max}$ (Fig. 6b).\par

\appendix

\section{Parameter Values}
 
To constrain the physiological parameters of our models, we used data from 
large-scale studies of the yeast transcriptome and proteome; see Table I. 
These data provide typical ranges and values for our parameters. 
First, we converted numbers of molecules into nanomolar (nM) 
concentrations using the cell volume $V=40$ ${\rm fL}$ for haploid yeast. 
For the ATC, we assumed that the basal mRNA transcription would be low. 
Thus, $\rho_f$ for the ATC was constrained to values from the bottom 5th 
percentile to the median of all mRNA synthesis rates \cite{Miller:2011fp}. 
Similarly, $\rho_f$ for the RTC was constrained to values from the median 
to top 95th percentile. In the case of the ATC, the constraint 
$\rho_f<\rho_0<\rho_b$ ensured that the inhibitor can completely 
titrate the constitutively-expressed activator 
when the inhibitor is maximally produced at $\rho_b$, but not when it is 
expressed at the basal rate $\rho_f$. Similarly, for the RTC, the constraint 
$\rho_b<\rho_0<\rho_f$ ensures that constitutively-expressed inhibitor can 
completely titrate the repressor when the repressor is produced at the 
repressed rate $\rho_b$, but not $\rho_f$. We set 
$\rho_0=\sqrt{\rho_f\rho_b}$ to satisfy both conditions. mRNA degradation 
rate ranged from the bottom 5th percentile to top 95th percentile values for 
all genes \cite{Miller:2011fp}. To obtain a rough approximation 
for the translation rates, we assumed a constitutive gene expression 
model for all genes:

\begin{align}
\frac{d[r]}{dt}&={\rho}-\delta_m[r]\\
\frac{d[P]}{dt}&={\beta}[r]-\delta_p[P]
\end{align}
At steady state, $\beta=\frac{[P]\delta_p\delta_m}{\rho}$. Protein 
concentrations and degradation rates were taken from 
\cite{Ghaemmaghami:2003ds,Belle:2006hv}. We calculated 
$\beta$ for all genes and used the median value in our model. We also 
assumed that our activators, repressors, and inhibitors are not 
actively degraded and are diluted by growth. Thus, 
$\delta_p=ln(2)/T$, where $T=90$ mins is the duration of the 
yeast cell cycle. The proteins in our models were based on a mammalian 
transcription factor basic leucine zipper (bZIP) protein C/EBP$\alpha$ and its 
dominant-negative inhibitor (3HF) \cite{Buchler:2009fz}. We used previously 
measured rates for protein-protein interaction kinetics \cite{Buchler:2009fz}. 
Since we did not know the DNA unbinding rate for C/EBP$\alpha$, we considered 
the range for the known DNA unbinding rates for other bZIP proteins 
\cite{GTC:GTC711,FEBS:FEBS1897,Okahata1998,Geertz09102012}. 
The thermodynamic dissociation constant ($K_d$) of C/EBP$\alpha$ to 
its specific DNA binding site is known \cite{Cao01091991}. We set the 
DNA association rate to $\alpha = \theta/K_d$. Finally, we varied the 
activation/repression strength $f$ from 1 to 30, to consider both strong 
and weak activation/repression.

\begin{table}
\caption{Parameter values\label{}}
\begin{ruledtabular}
\begin{tabular}{@{\vrule height 10.5pt depth4pt width0pt}cccc}
Parameter&Min&Max&Reference\\
\hline

$\theta({\rm min}^{-1})$&0.0188&34.5
&\cite{GTC:GTC711,FEBS:FEBS1897,Okahata1998,Geertz09102012}\\
ATC $\rho_f($${\rm min}^{-1})$&0.0509&0.1781&\cite{Miller:2011fp}\\
RTC $\rho_f($${\rm min}^{-1})$&0.1781&0.8928&\cite{Miller:2011fp}\\
$\delta_m({\rm min}^{-1})$&0.0159&0.1516&\cite{Miller:2011fp}\\
$f$&1&30\\
ATC $\rho_b($${\rm min}^{-1})$&\multicolumn2c{$f \cdot \rho_f$}\\
RTC $\rho_b($${\rm min}^{-1})$&\multicolumn2c{$\rho_f / f$}\\
$\alpha({\rm nM}^{-1}$ ${\rm min}^{-1})$&\multicolumn2c
{$\theta/3.344$ ${\rm nM}$}&\cite{Cao01091991}\\
$\rho_0($${\rm min}^{-1})$&\multicolumn2c{$\sqrt{\rho_f\rho_b}$}\\
$\beta({\rm min}^{-1})$&\multicolumn2c{14.1}&
\cite{Miller:2011fp,Ghaemmaghami:2003ds,Belle:2006hv}\\
$\delta_p({\rm min}^{-1})$&\multicolumn2c{0.0077}\\
$\gamma({\rm nM}$ ${\rm min}^{-1})$&\multicolumn2c{0.6}&
\cite{Buchler:2009fz}\\
$\epsilon_1({\rm min}^{-1})$&\multicolumn2c{6}&\cite{Buchler:2009fz}\\
$\epsilon_2({\rm min}^{-1})$&\multicolumn2c{0.024}&\cite{Buchler:2009fz}\\
$V({\rm fL})$&\multicolumn2c{40}&\cite{Buchler:2008bh}\\
\end{tabular}
\end{ruledtabular}
\end{table}

\section{Methods}
We scanned the parameter space for oscillations by running simulations on 
MATLAB (Mathworks) using ode15s
for 2000 min and recording the minima and maxima of the activator/repressor 
homodimer during the last 1000 min. We imposed the following restrictions: 1) 
the last minimum should be below $K_d$ so that DNA-binding is not saturated, 
2) the last maximum should be greater than $2K_d$ so that the change in 
transcription is noticeably altered. While this restriction slightly 
underestimates the number oscillatory solutions, it ensures that a 
synthetic version of these circuits would produce detectable oscillations. 
We verified that our definition gave similar results to a less stringent 
criterion for oscillation.\par
We used the direct Gillespie method to perform the stochastic simulations 
\cite{Hoops:2006ui}. We ran the simulations for $10^6$ min and recorded the 
concentration of the regulated mRNA (inhibitor for the ATC and the repressor 
for the RTC). We then normalized the concentration such that the average 
would be zero and evaluated the autocorrelation function. We then fit the 
function $C(t)=e^{-t/\tau_0}\cdot{\rm cos}(2{\pi}t/\tau)$ to the first 1500 min of 
the autocorrelation function to measure the decay constant $\tau_0$ and 
period $\tau$. The ratio $\tau_0/\tau$ describes how rapidly the envelope 
of autocorrelation function decays per oscillation period.

\begin{acknowledgments}
We thank Joshua Socolar for helpful comments. This work was funded by 
an NIH Director's New Innovator Award (DP2 OD008654-01) and 
Burroughs Wellcome Fund CASI Award (BWF 1005769.01). An SBML version of the ATC and RTC oscillators can be downloaded from the BioModels Database (MODEL1512100000 and MODEL1512100001, respectively).
\end{acknowledgments}

\end{document}


\title{Supplementary Material}
\author{Sargis Karapetyan}
\affiliation{Department of Physics, Duke University, Durham, NC 27708}
\affiliation{Center for Genomic \& Computational Biology, Durham, NC 27710}

\author{Nicolas E. Buchler}
\affiliation{Department of Physics, Duke University, Durham, NC 27708}
\affiliation{Center for Genomic \& Computational Biology, Durham, NC 27710}
\affiliation{Department of Biology, Duke University, Durham, NC 27708}
\pacs{}
\maketitle

\section{Promoter with multiple binding sites}

We first derive the chemical kinetic equations for 2 binding sites and then generalize to $n$ binding sites \cite{Lengyel2014}.  Similar to previous work \cite{Bintu:2005bn}, we calculate the promoter sensitivity of activators and repressors in the limit of thermodynamic equilibrium.  

\subsection{Promoter with 2 equal binding sites}

We first begin with an explicit derivation that includes each binding state combination:

\begin{align}
\frac{d[G_{00}]}{dt} &= - \alpha \cdot [G_{00}] [X_2] + \theta \cdot [G_{10}] -\alpha \cdot [G_{00}] [X_2] + \theta \cdot [G_{01}] \\
\frac{d[G_{01}]}{dt} &= \alpha \cdot [G_{00}] [X_2] - \theta \cdot [G_{01}] -\alpha \cdot [G_{01}] [X_2] + \theta \cdot [G_{11}] \\
\frac{d[G_{10}]}{dt} &= \alpha \cdot [G_{00}] [X_2] - \theta \cdot [G_{10}] -\alpha \cdot [G_{10}] [X_2] + \theta \cdot [G_{11}] \\
\frac{d[G_{11}]}{dt} &= \alpha \cdot [G_{10}] [X_2] - \theta \cdot [G_{11}]  + \alpha \cdot [G_{01}] [X_2] - \theta \cdot [G_{11}]
\end{align}

where $[G_T] = [G_{00}] + [G_{01}] + [G_{10}] + [G_{11}]$ and $[X_2]$ is the homodimer transcription factor concentration.  We rewrite these equations in a compact form:  

\begin{align}
\frac{d[G_{0}]}{dt} &= - 2 \alpha \cdot [G_{0}] [X_2] + \theta [G_{1}]  \\
\frac{d[G_{1}]}{dt} &= 2 \alpha \cdot [G_{0}] [X_2] - \theta [G_{1}] -\alpha [G_{1}] [X_2] + 2 \theta [G_{2}] \\
\frac{d[G_{2}]}{dt} &= \alpha \cdot [G_{1}] [X_2] - 2 \theta [G_{2}]  
\end{align}

where $[G_0] = [G_{00}]$, $[G_1] = [G_{01}] + [G_{10}]$, and $[G_2] = [G_{11}]$. This simplification is valid when all binding sites have equivalent kinetics and the regulator $X_2$ binds independently to each  site. 

\subsubsection{Promoter synthesis rate at equilibrium}

At equilibrium, Eqs. (5-7) are equal to 0. Thus,  

\begin{align}
2  [G_0] [X_2] &= K_d \cdot [G_1]  \\
(K_d + [X_2]) [G_1] &= 2 [G_0] [X_2]  + 2 K_d \cdot [G_2] \\
{[}G_1] \cdot [X_2] &= 2 K_d \cdot [G_2] 
\end{align}

where $K_d = \theta / \alpha$.  Solving these equations with constraint $[G_T] = [G_{0}] + [G_{1}] + [G_{2}]$: 

\begin{align}
[G_0 ] = [G_T] \cdot \frac{1}{(1 + \frac{[X_2]}{K_d})^2} \\
{[}G_1 ] =  [G_T] \cdot \frac{2 \cdot (\frac{[X_2]}{K_d})}{(1 + \frac{[X_2]}{K_d})^2} \\
{[}G_2 ] = [G_T] \cdot \frac{(\frac{[X_2]}{K_d})^2}{(1 + \frac{[X_2]}{K_d})^2}
\end{align}

The promoter synthesis rate $P_2$ where any bound state has the same gene expression ($\rho_b$) and the unbound state has gene expression ($\rho_f$) is given by: 

\begin{equation}
P_2 (X_2) = [G_T] \cdot \left[ \frac{\rho_b \left[2 \cdot X_2 + (X_2)^2 \right] + \rho_f }{(1+X_2)^2} \right] = [G_T] \cdot \left[ \rho_b + \frac{(\rho_f - \rho_b) }{(1+X_2)^2} \right]
\end{equation}
where $X_2 = \frac{[X_2]}{K_d}$.

\subsection{Promoter with n equal binding sites}
By generalizing our previous derivation for an arbitrary number of binding sites, we get $n+1$ differential equations, where $n$ is the total number of binding sites and the index $i$ spans $1 \le i \le n-1$. 

\begin{align}
\frac{d[G_0]}{dt} &= -n \alpha \cdot [G_0] [X_2] + \theta [G_1] \\
&\vdots& \notag \\
\frac{d[G_i]}{dt} &= (n-i+1) \alpha \cdot [G_{i-1}] [X_2] - (i \cdot \theta + (n-i) \alpha \cdot [X_2]) [G_{i}] + (i+1) \theta \cdot [G_{i+1}] \\
&\vdots& \notag \\
\frac{d[G_n]}{dt} &=  \alpha \cdot [G_{n-1}] [X_2] - n \theta \cdot [G_n] 
\end{align}

$G_i$ denotes the promoter states with $i$ out of total $n$ binding sites occupied by activator (X=A) or repressor (X=R) dimers. The factors in front of each term represents the amount of degeneracy of each state, i.e $[G_i]$ has $i$ bound sites, thus $i$ ways of switching to $[G_{i-1}]$. Therefore, we have the term $i\cdot \theta [G_i]$. At the same time, $[G_i]$ has $n-i$ vacant sites, so it has $n-i$ ways of switching to the state $[G_{i+1}]$. Thus, we have the term $(n-i) \alpha \cdot [X_2] [G_{i}]$.  Note that our model assumes that the binding of a transcription factor does not affect the binding or unbinding of the next transcription factor to an adjacent site (no cooperativity).  We also assume that transcription is fully activated/repressed for any bound state of the promoter, i.e. transcription is equal to $\rho_b$ for $G_1$, $G_2$, $\ldots$, $G_n$. Although not shown, the ODE for the rate of change in dimer concentration Eq. (1f) in the main text is also modified to reflect binding and unbinding from each state $G_i$.\par

\subsubsection{Promoter synthesis rate at equilibrium}

Setting Equations (15-17) equal to zero and solving with the constraint $[G_T] = [G_{0}] + [G_{1}] + \ldots + [G_{n}]$:

\begin{align}
[G_0 ] &= [G_T] \cdot \frac{1}{(1 + \frac{[X_2]}{K_d})^N} \\
&\vdots& \notag  \\
{[}G_i ] &= [G_T] \cdot \frac{\frac{N!}{i!(N-i)!} \cdot (\frac{[X_2]}{K_d})^i}{(1 + \frac{[X_2]}{K_d})^N} \\
&\vdots& \notag \\
{[}G_N ] &= [G_T] \cdot \frac{(\frac{[X_2]}{K_d})^N}{(1 + \frac{[X_2]}{K_d})^N}
\end{align}

The promoter synthesis rate $P_n$ is thus given by: 

\begin{equation}
P_n (X_2) = [G_T] \cdot \left[ \rho_b + \frac{(\rho_f - \rho_b) }{(1 + X_2)^n} \right]
\end{equation}

where $X_2 = \frac{[X_2]}{K_d}$.

\subsection{Logarithmic sensitivity analysis}

We compute the logarithmic sensitivity of the promoter: $S = \frac{{\rm dlog}P}{{\rm dlog}X_2} = \frac{X_2}{P(X_2)} \cdot \frac{dP}{dX_2}$: 
 
 \begin{equation}
S(X_2) = \frac{n (\rho_b - \rho_f) X_2}{(1+X_2) \cdot (\rho_f + \rho_b [ (1+X_2)^n - 1] )}
\end{equation}

\subsubsection{Activation}

We consider the simplified case of ideal activation, where $\rho_f = 0$.  Thus, 

 \begin{equation}
S(X_2) = \frac{n X_2}{(1+X_2) \cdot [ (1+X_2)^n - 1] }
\end{equation}

$S$ is a monotonically decreasing function of $X_2$ with boundaries $S(0) = 1$ and $S(\infty)=0$.  Thus, activation is never ultrasensitive (i.e. $|S| \le 1$ for any value of $X_2$, $n$, or $\rho_b$). Additionally, $|S|$ is also a monotonically decreasing function of $n$ for all $X_2 >0$; see Figure S1. The boundaries are at $|S(0)|=\frac{X_2}{(1+X_2){\cdot}ln(1+X_2)}$ and $|S(\infty)|=0$. This explains why the ATC oscillator phase space decreases with the addition of multiple DNA binding sites.

\subsubsection{Repression}

We consider the case of ideal repression\, where $\rho_b = 0$.  Thus, 

 \begin{equation}
S(X_2) = \frac{-n X_2}{(1+X_2)}
\end{equation}

$S$ is a monotonically decreasing function of $R$ with boundaries $S(0) = 0$ and $S(\infty)=-n$.  Promoter repression is ultrasensitive (i.e. $|S| > 1$) for $X_2>\frac{1}{n-1}$ and $n>1$; see Figure S1. $|S|$ is a monotonically increasing function of $n$ for all $X_2 >0$. The boundaries are at $|S(0)| = 0$ and $|S(\infty)|=\infty$. Thus, the RTC oscillator phase space increases with the addition of multiple DNA binding site.

\section{Kinetics of complete unbinding from multiple sites}

The probability of unbinding for a single binding site is given by a Poisson distribution: 

\begin{equation}
p(t) = \theta e^{-\theta t}
\end{equation}

where $\theta$ is the unbinding rate.  The average unbinding time for a single site is given by $<\tau> = \int_0^{\infty} t \cdot p(t) dt$ and is equal to $\frac{1}{\theta}$.  Consider the case where all $n$ binding sites are filled and where the concentration of regulator abruptly changes to 0 (such that no rebinding occurs). We calculate the probability distribution of time for {\it complete} unbinding from all $n$ binding sites. The probability that the slowest unbinding event is $\tau$ and the other $n-1$ events are less than $\tau$ is: 

\begin{align}
{\cal P}(\tau | n, \theta) &= \left[ \int_0^{\tau} p(t) dt \right]^{n-1} \cdot p(\tau) \\
{\cal P}(\tau | n, \theta) &= \left[ 1- e^{-\theta \tau} \right]^{n-1} \cdot \theta e^{-\theta \tau}
\end{align}

When we normalize the distribution ($\int_0^{\infty} {\cal P}(\tau | n, \theta) dt = 1$): 

\begin{equation}
{\cal P}(\tau | n, \theta) = n \left[ 1- e^{-\theta \tau} \right]^{n-1} \cdot \theta e^{-\theta \tau}
\end{equation}

The average $<\tau>$ can be calculated by $\int_0^{\infty} \tau \cdot {\cal P}(\tau) d\tau$: 

\begin{equation}
<\tau> = \frac{H_n}{\theta} = \frac{1}{\theta_n}
\end{equation}

where $H_n = 1 + \frac{1}{2} + \frac{1}{3} + \ldots + \frac{1}{n}$ (Harmonic number) and $\theta_n$ is the effective DNA unbinding rate. 
\pagebreak
\section{Supporting figures}

\begin{figure}[!h]
\includegraphics[width=.45\textwidth]{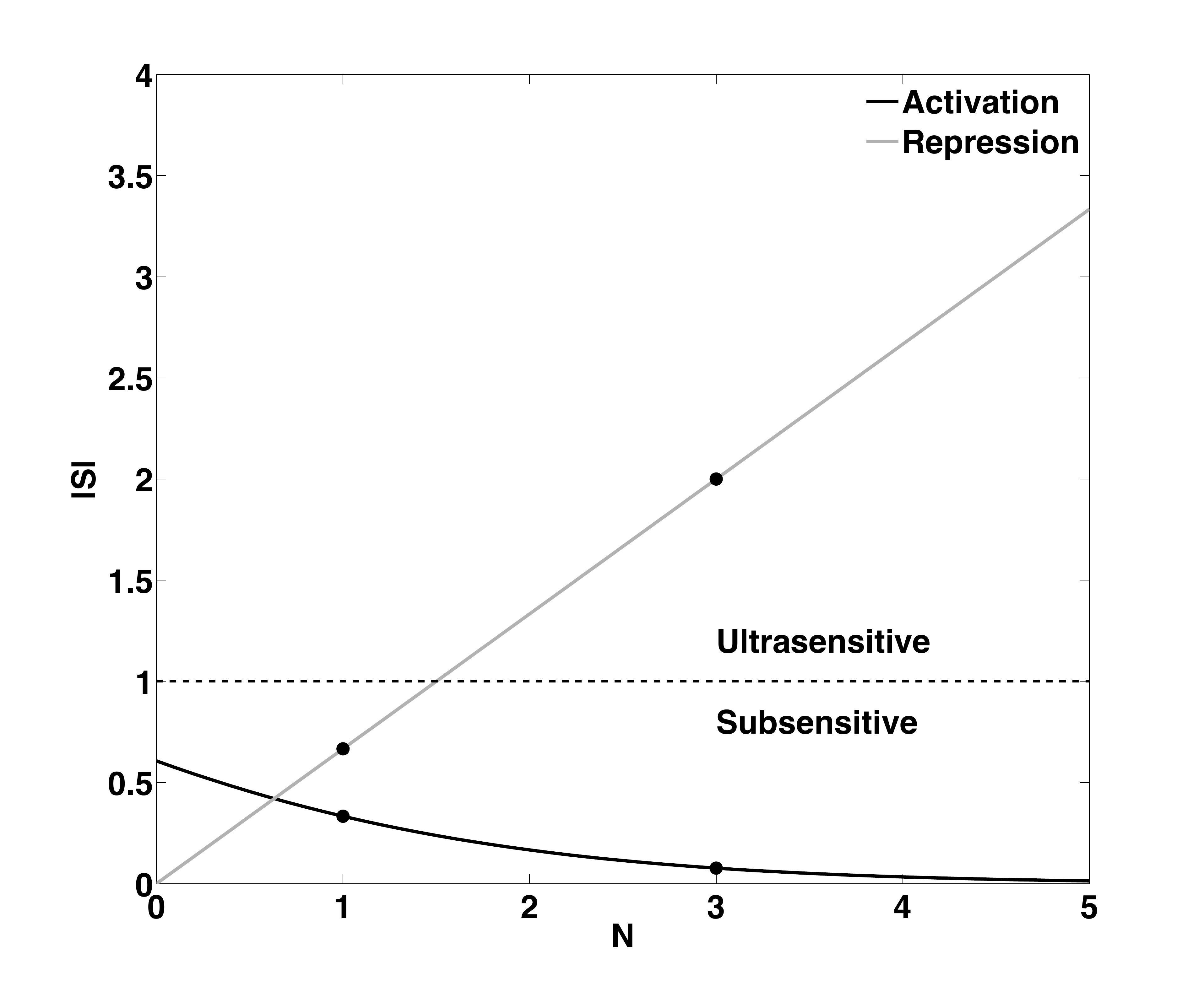}
\caption{Multiple binding sites affect activation- and repression-based promoters differently. The absolute value of logarithmic sensitivity $S$ is plotted as a function of number of binding sites $N$, with $X_2=2$. The points of 1 and 3 binding sites considered in this study are marked with circles.}
\end{figure}

\begin{figure}[!h]
\includegraphics[width=.5\textwidth]{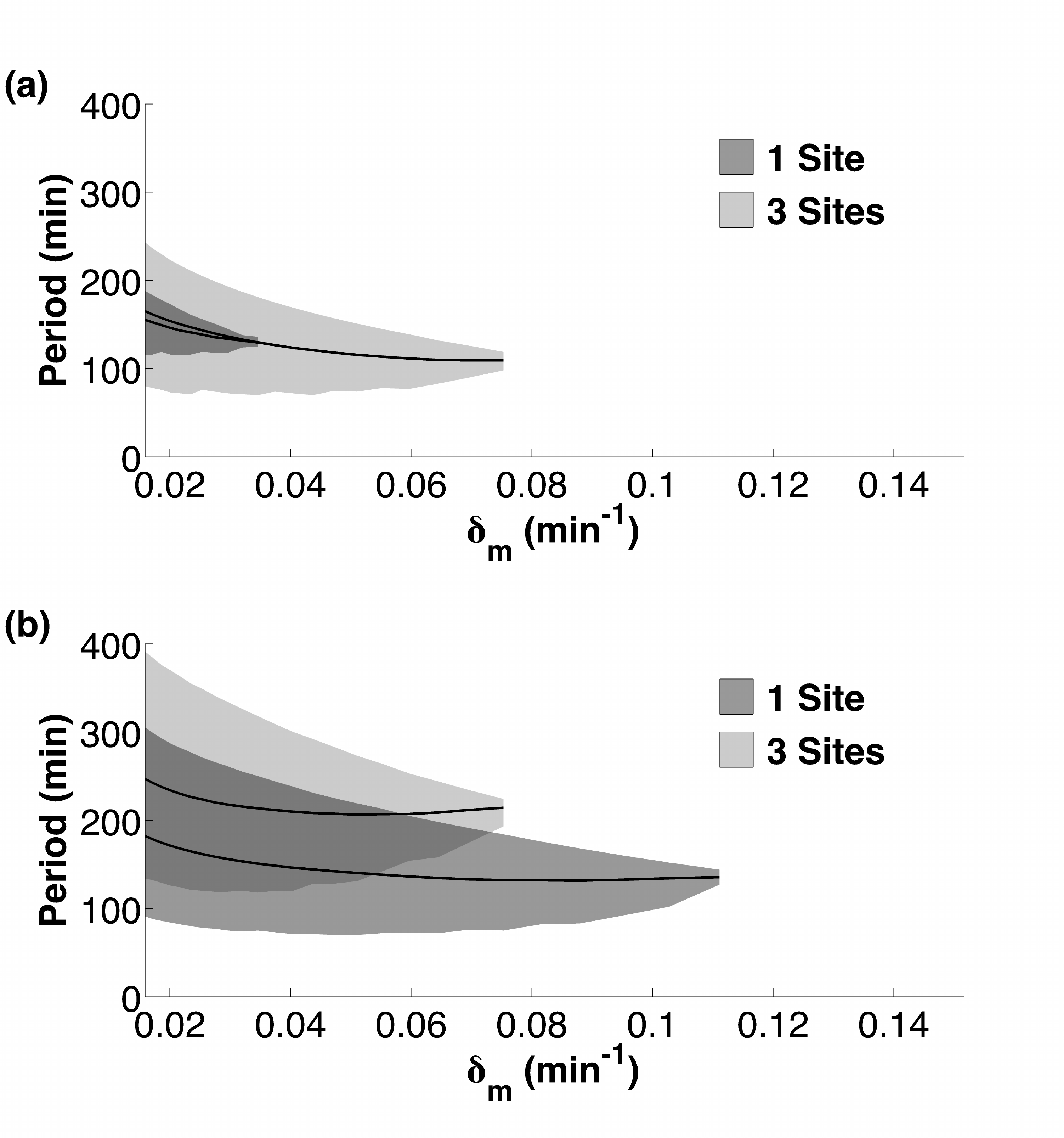}
\caption{The mRNA degradation rate $\delta_m$ sets the period of the oscillations for RTC (a) and ATC (b) at slow mRNA degradation rates. The mean period of oscillatory solutions for a given $\delta_m$ is shown (solid black line) with the shaded area representing the range of periods.}
\end{figure}